\documentclass[11pt, a4paper]{article}
\usepackage{moriond,epsfig}

\def\be{\begin{equation}}
\def\ee{\end{equation}}
\def\bea{\begin{eqnarray}}
\def\eea{\end{eqnarray}}

\newcommand{\fs}{\, .}
\newcommand{\co}{ \, ,}

\newcommand{\ChPT}{\raisebox{0.16em}{$\chi$}PT}
\begin{document}
\vspace*{4cm}
\title{INSIGHTS AND PUZZLES IN LIGHT QUARK PHYSICS}
\author{H. Leutwyler}
\address{Institute for theoretical physics,
University of Bern\\ Sidlerstr. 5, CH-3012 Bern, Switzerland}

\maketitle\abstracts{Recent work in light flavour hadron physics is reviewed.
  In particular, I discuss the significance of the progress achieved with
  light dynamical quarks on the lattice for the effective low energy theory of
  QCD. Also, I draw attention to some puzzling results from NA48 and KTeV
  concerning the scalar form factor relevant for $K_{\mu3}$ decay -- taken at
  face value, these indicate physics beyond the Standard Model.}
\noindent
{\small¥{\it Keywords}: Quantum chromodynamics, light quarks}

\section{Introduction}
At low energies, the most important characteristic of QCD is that the energy
gap is very small -- a consequence of the fact that the $u$- and $d$-quarks
happen to be very light. In the theoretical limit where the quark masses $m_u$
and $m_d$ are set equal to zero, the Hamiltonian of QCD acquires an exact,
spontaneously broken symmetry. The spectrum of the theory does then not have
an energy gap at all: in the limit, the pions become massless particles,
playing the role of the Goldstone bosons which necessarily occur if a
continuous symmetry spontaneously breaks down.

The remarkable theo\-re\-ti\-cal progress made in light flavour hadron physics
in recent years relies on the fact that these properties can be used to
construct an effective theory (``chiral perturbation theory'', referred to as
\ChPT), that allows us to analyze the low energy structure of QCD in a
controlled manner. In this framework, all of the Green functions formed with
the quark currents can be calculated in terms of the coupling constants
occurring in the effective Lagrangian, order by order in the chiral expansion.
For recent reviews of \ChPT, I refer to Scherer$\;$\cite{Scherer},
Bijnens \cite{Bijnens} and Colangelo.\cite{Colangelo Erice} An up-to-date
account of our knowledge of the effective coupling constants can be found in a
recent conference report by Ecker.\cite{Ecker}

In my talk, I focused on a few selected aspects of this development. In
particular, I emphasized the fact that the progress achieved in the numerical
simulation of QCD on a lattice made it possible to reach sufficiently light
quarks, so that the extrapolation to the quark masses of physical interest can
be done in a controlled manner, using \ChPT. I will discuss
this in some detail below. Another recent development which I will briefly
report on, concerns the semileptonic decay $K\rightarrow\pi\mu\nu$. Both KTeV
and NA48 have recently published new results on the scalar form factor of this
decay, which are in flat conflict with a venerable low energy theorem,
established by Callan and Treiman in the sixties.\cite{Callan Treiman} If these
results are confirmed, then the Standard Model is in conflict with observation
in one of those reactions which we thought are best understood.

For lack of space, I cannot cover the third topic which I dealt with at these
Rencontres: the progress made in understanding the properties of the
interaction among the pions. This interaction plays a crucial role in many
contexts -- the Standard Model prediction for the magnetic moment of the muon
is perhaps the most prominent example. To close this introductory section, I
briefly list the corresponding keywords and indicate where more information
about this can be found.  The $\pi\pi$ scattering amplitude has been
calculated to NNL in \ChPT.\cite{BCEGS} The resulting representation is very
accurate in the interior of the Mandelstam triangle, but in the physical
region, the convergence of the series is rather slow. The range of energies
where the chiral representation yields meaningful results can be extended with
the inverse amplitude trick, but one is then leaving the territory where model
independent statements can be made. There is a general method that does not
suffer from such shortcomings: in a limited region of the complex plane,
dispersion theory imposes a set of exact relations between the real and
imaginary parts of the partial wave amplitudes, the Roy
equations.\cite{Roy,CCL} The region where these equations are valid includes
the poles on the second sheet generated by the lowest resonances of QCD:
$\sigma$, $\rho$, $\omega$, $f_0(980)$, $a_0(980)$. The crucial parameters
that control the low energy properties of the scattering amplitude in this
framework are the two subtraction constants.  It is convenient to identify
these with the two S-wave scattering lengths, because the low energy theorems
of \ChPT$\;$make very sharp predictions for these. Together with the low
energy theorems, the Roy equations pin the scattering amplitude down within
remarkably small uncertainties.\cite{CGL} The angular momentum barrier ensures
that the S- and P-waves dominate at low energies, but the framework also
yields very accurate predictions for partial waves with higher angular
momenta. On the basis of this method, it can be demonstrated beyond reasonable
doubt that the lowest resonance of QCD carries the quantum numbers of the
vacuum and the position of the corresponding pole can be worked out rather
accurately.\cite{CCL} For an overview of these developments, I refer to the
proceedings of Chiral Dynamics 2006 and the references quoted therein.\cite{CD
  2006}
\section{Size of the energy gap}

In order to illustrate the progress achieved on the lattice, I consider one of
the key issues in QCD: understanding the size of the energy gap. The quark
masses $m_u$ and $m_d$ are very small, but they are different from zero.
Accordingly, the Hamiltonian of QCD is not invariant under chiral rotations --
the quark mass term breaks chiral symmetry and there is an energy gap: the
symmetry breaking equips the Goldstone bosons with a mass. The quark masses
$m_u,m_d$ represent a quantitative measure of the strength of the symmetry
breaking. We know that the symmetry breaking is very small, but,
unfortunately, the Standard Model does not offer an understanding of why that
is so -- the entire fermion mass pattern looks bizarre and yet remains to be
understood.

\begin{figure}[thb]
  \includegraphics[width=8.2cm]{mpi.eps}\hspace{0.5cm}
  \includegraphics[width=6cm]{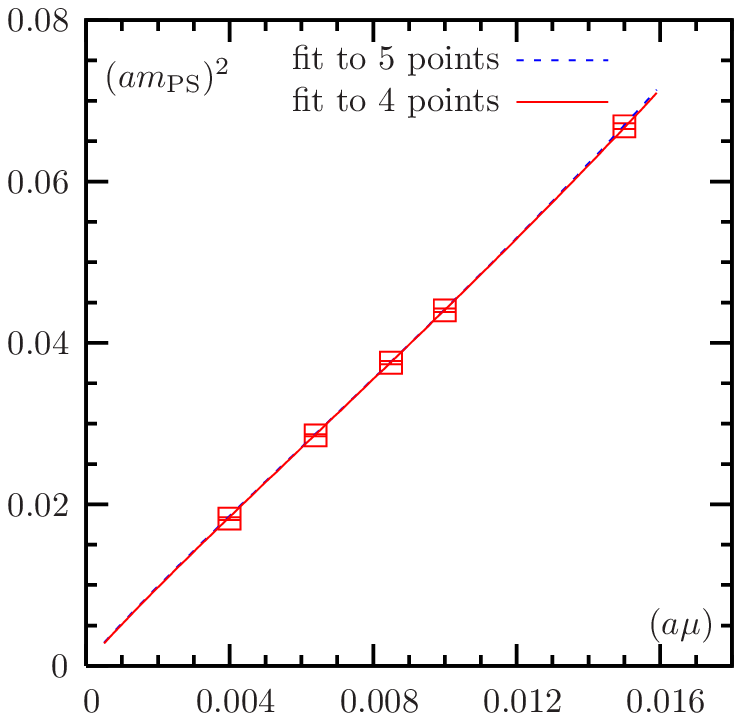}\\

  {\footnotesize \rule{2.5cm}{0cm}M.\ L\"uscher, Lattice 2005 \cite{Luscher}
    \hspace{2.7cm} Ph.~Boucaud {\it et al.}  [ETM Collaboration] \cite{ETM}}
\caption{Lattice results for $M_\pi^2$ as a function of the quark mass}
\end{figure}

As pointed out by Gell-Mann, Oakes and Renner,\cite{GMOR} the square of the
pion mass is proportional to the strength of the symmetry breaking,
$M_\pi^2\propto (m_u+m_d)$. This property can now be checked on the lattice,
where -- in principle -- the quark masses can be varied at will. Fig.\ 1 shows
the result for two recent lattice simulations of QCD with two flavours. In
view of the fact that in these calculations, the quarks are treated
dynamically, the quality of the data is impressive. The masses are
sufficiently light for \ChPT$\;$to allow a meaningful extrapolation to the
quark mass values of physical interest.  The results indicate that the ratio
$M_\pi^2/(m_u+m_d)$ is nearly constant out to values of $m_u, m_d$ that are
about an order of magnitude larger than in nature.

\section{\boldmath Lattice determinations of the effective coupling constants
  $\bar{\ell}_3$ and $\bar{\ell}_4$} The Gell-Mann-Oakes-Renner relation
represents the leading term in the expansion in powers of the quark masses. At
next-to-leading order, this expansion contains a logarithm: \be\label{Mpi one
  loop} M_\pi^2= M^2\left\{1 +\!\frac{M^2}{32\pi^2 F^2}\, \ln
  \frac{M^2}{\Lambda_3^2}\!+\!O(M^4)\right\}\co\hspace{1cm} M^2\equiv
B(m_u+m_d)\fs \ee Chiral symmetry fixes the coefficient of the logarithm in
terms of the pion decay constant $F$, but does not determine the scale
$\Lambda_3$ of the logarithm. A crude estimate was obtained more than 20 years
ago,\cite{GL 1984} on the basis of the SU(3) mass formulae for the
pseudoscalar octet: \be\label{l3bar}
0.18\,\mbox{GeV}<\Lambda_3<2\,\mbox{GeV}\hspace{1cm}\Longleftrightarrow
\hspace{1cm}\bar{\ell}_3\equiv \ln \frac{\Lambda_3^2}{M_\pi^2} = 2.9\pm
2.4\fs\nonumber \ee The logarithmic term implies that the lines in Fig.\ 1
cannot be straight. For the central value, $\bar{\ell}_3 = 2.9$, the formula
(\ref{Mpi one loop}) yields only little curvature, but if $\bar{\ell}_3$ was
at the lower (upper) end of the quoted range, the plot of $M_\pi^2$ versus
$m_u=m_d$ would be strongly bent upwards (downwards), visibly departing from
the lattice results shown in Fig.\ 1, already at the lowest quark mass value.
Evidently, these results strongly constrain the value of the coupling constant
$\bar{\ell}_3$.

\begin{table}[thb]
\caption{Determinations of the effective coupling constants
  $\bar{\ell}_3$ and $\bar{\ell}_4$}\centering
\begin{tabular}{|c|c|c|c|c|}\hline \rule{0.5cm}{0cm}&\rule{2.9cm}{0cm} &
\rule{2.9cm}{0cm} & \rule{2.9cm}{0cm}&\rule{2.9cm}{0cm}\\
&\ChPT$\;\,$\cite{CGL,GL 1984}&MILC \cite{Bernard CD2006}& Del Debbio et
al.\cite{Del Debbio et al}&ETM \cite{ETM}\\ 
\hline\rule{0em}{1.5em}
$\bar{\ell}_3$&$2.9\pm 2.4$ &$0.6\pm 1.2$&$3.0\pm 0.5$&
$3.62\pm 0.12$\\
\rule{0.4em}{0em}$\bar{\ell}_4$&$4.4\pm 0.2$ &$3.9\pm 0.5$& {\bf ---}&
$4.52\pm 0.06$\\
& & & &  \\ \hline
\end{tabular}
\end{table}
The first row in table 1 compares the number in equation (\ref{l3bar}) with
recent determinations of $\bar{\ell}_3$ on the lattice. The second row lists
the analogous results for the coupling constant $\bar{\ell}_4$, which controls
the quark mass dependence of the pion decay constant. To my know\-ledge, the
first lattice calculation of effective coupling constants based on dyna\-mical
quarks was carried out by the MILC collaboration.\cite{MILC} In that project,
the coupling constants $L_4,L_5,L_6,L_8$, which occur in the effective chiral
SU(3)$\times$SU(3) Lagrangian at first nonleading order, were determined by
analyzing the quark mass dependence of $M_\pi,M_K,F_\pi$ and $F_K$ by means of
\ChPT. The corresponding values of $\bar{\ell}_3$ and $\bar{\ell}_4$ are
readily worked out, using standard one loop formulae.\cite{GL SU3} The numbers
quoted in the third column are obtained from a recent update of the MILC
analysis, which relies on staggered quarks.\cite{Bernard CD2006} The results
obtained by Del Debbio et al.\cite{Del Debbio et al} are based on two flavours
of Wilson quarks, while the European Twisted Mass Collaboration \cite{ETM}
uses two flavours of twisted mass Wilson quarks. All of the numbers are
consistent with the values found in \ChPT, but some of them are more accurate.
Unfortunately, the results obtained with staggered quarks do not agree well
with those based on Wilson quarks.  Possibly, this is related to the fact that
the latter calculations treat all quarks except $u$ and $d$ as infinitely
heavy -- as emphasized by Stern and collaborators, the strange quark may play
a significant role here.\cite{Descotes et al 2004} Lattice results for 3
flavours of light Wilson quarks should become available soon, so that it
should be possible to identify the origin of the difference.

\begin{figure}[thb]\centering
\includegraphics[width=7.8cm,angle=-90]{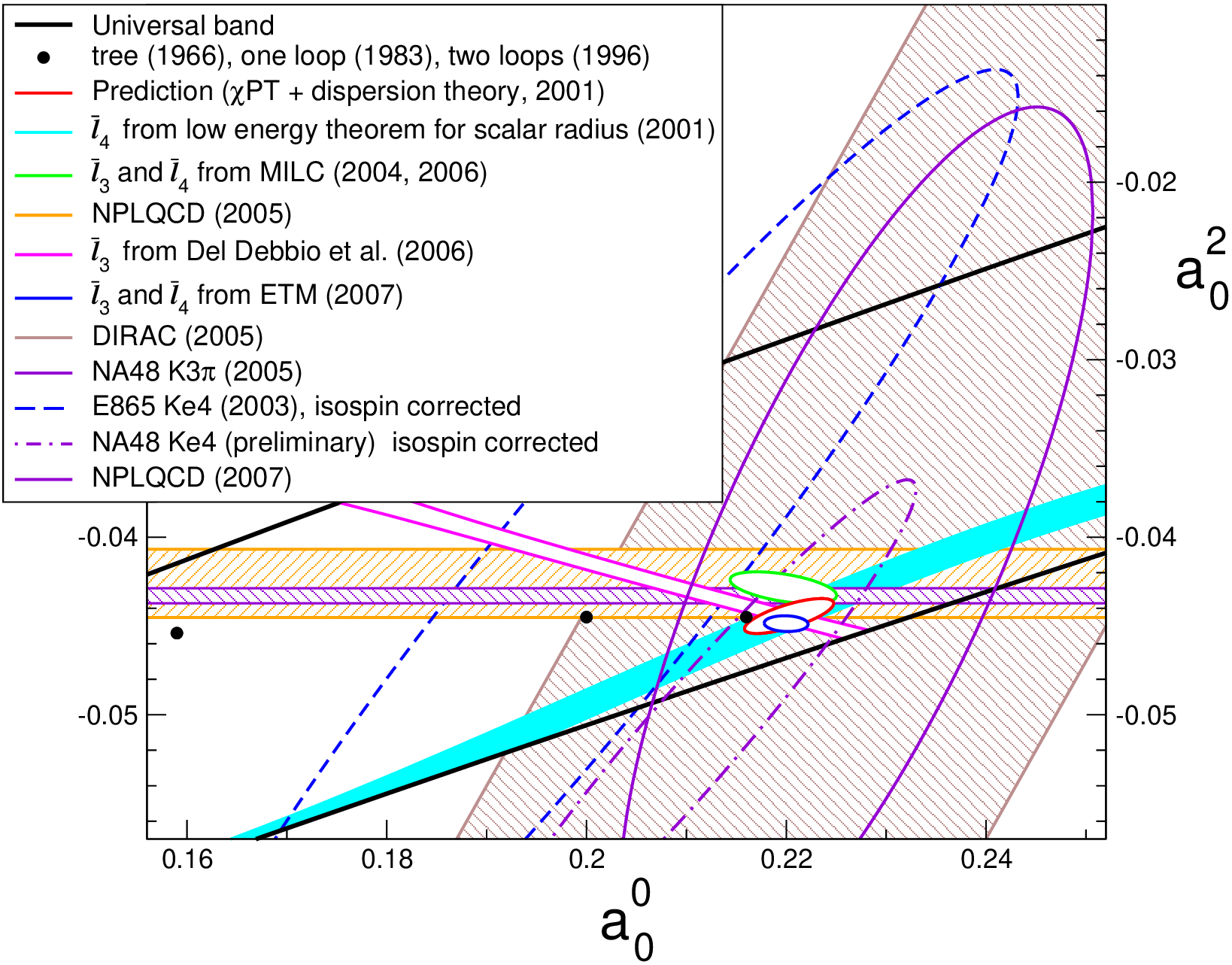}
\caption{\label{figa0a2theory}Theoretical and experimental results for
  the $\pi\pi$ S-wave scattering lengths.}
\end{figure}  

The values of the low energy constants $\bar{\ell}_3$ and $\bar{\ell}_4$ also
enter the theoretical prediction for the $\pi\pi$ scattering lengths. This is
illustrated in Fig.\ 2, where theory is compared with experiment (the lattice
results for $\bar{\ell}_3,\bar{\ell}_4$ are converted into corresponding
values for $a_0^0,a_0^2$ using \ChPT$\;$and dispersion theory \cite{CGL,CD
  2006}).  While the lattice results, DIRAC and the NA48 data on the cusp in
$K\rightarrow 3\pi$ confirm our predictions for $a_0^0,a_0^2$, the $K_{e4}$
data of NA48/2 give rise to a puzzle: the phase extracted from the transition
amplitude deviates from the theoretical prediction for
$\delta_0^0-\delta_1^1$. The discrepancy originates in the fact that neutral
kaons may first decay into a pair of neutral pions, which then undergoes
scattering and winds up as a charged pair. As pointed out by Colangelo, Gasser
and Rusetsky, the mass difference between the charged and neutral pions
affects this process in a pronounced manner: it pushes the phase of the
transition amplitude up by about half a degree -- an isospin breaking effect,
due almost exclusively to the electromagnetic interaction. The dash-dotted
line in Fig.\ 2, which is taken from the talk of B.\ Bloch-Devaux at KAON
2007, shows the likelihood contour ($\chi^2=\chi_{min}^2+2.3$) of the so
corrected, preliminary NA48/2 data. The intersection with the region allowed
by the low energy theorem for the scalar radius yields $a_0^0=0.220(9)$.
However, there is a discrepancy with the E865 data, for which the likelihood
contour is shown as a dashed line (kindly provided by G.  Colangelo): the
isospin correction spoils the good agreement between these data and the
prediction. The discrepancy only concerns the region $M_{\pi\pi}>$ 350 MeV.
While E865 collects all events in this region in a single bin, the resolution
of NA48/2 is better. The fit to all $K_{e4}$ data is therefore dominated by
NA48/2. For a detailed discussion of these issues, I refer to the talks by B.\ 
Bloch-Devaux, G.\ Colangelo and J.\ Gasser at KAON 2007. I conclude that the
puzzle is gone: $K_{e4}$ confirms the theory to remarkable precision.

\section{\boldmath Puzzling results in $K_{\mu3}$ decay}
The low energy theorem of Callan and Treiman \cite{Callan Treiman} predicts
the size of the scalar form factor of the decay $K\rightarrow \pi\mu\nu$ at
one particular value of the momentum transfer, namely $t=M_K^2-M_\pi^2$:
\be\label{CT} f_0(M_K^2-M_\pi^2)=\frac{F_K}{F_\pi}+O(m_u,m_d) \fs\ee Within
QCD, the relation becomes exact if the quark masses $m_u$ and $m_d$ are set
equal to zero. The corrections of first nonleading order, which have been
evaluated long ago,\cite{GL form factors} are tiny: they lower the right hand 
side by $3.5\times 10^{-3}$. In the meantime, the chiral perturbation series of
$f_0(t)$ has been worked out to NNL.\cite{Post et al} As pointed out by Jamin,
Oller and Pich,\cite{Jamin Oller and Pich} the curvature of the form factor
can be calculated with dispersion theory, so that the prediction for the value
at $t=M_K^2- M_\pi^2$ can be converted into a rather accurate prediction for
the slope: $\lambda_0 = 0.0157(10)$. The dispersive representations of Jamin
et al.\cite{Jamin Oller and Pich} and Bernard et al.\cite{Stern Kmu3} agree
very well: theory reliably determines the curvature of the form factor.

Very recently, the NA48 collaboration published their analysis of the
$K^0_{\mu3}$ form factors.\cite{NA48 Kmu3} Their result for the scalar slope,
$\lambda_0=0.0117(7)(1)$, is in flat conflict with the prediction of Jamin,
Oller and Pich. Using the parametrization proposed by Bernard et
al.,\cite{Stern Kmu3} NA48 obtains $f_0(M_K^2-M_\pi^2)= 1.155(8)(13)\times
f_{+}(0)$. The value of $F_K/F_\pi$ is sensitive to the theoretical input used
for $f_+(0)$. Accurately measured branching ratios \cite{Palutan} and the
value of $V_{ud}$ (which has been determined to high precision on the basis of
nuclear $\beta$ decays) imply $F_K/F_\pi =1.241(4) \times f_+(0)$.  So, if the
NA48 results are correct, then the second term on the right hand side of
(\ref{CT}) must amount to a contribution of $0.086(17) \times f_+(0)$.  For a
correction of $O(m_u,m_d)$, this size is unheard of (as mentioned above, the
one loop approximation for this term amounts to -0.0035 and is thus smaller by
a factor of about 20). At the current experimental accuracy, the radiative
corrections may play a significant role. For $K_{\ell3}$ decay, these are
known to one loop of \ChPT,\cite{Neufeld} but the data yet remain to be
analyzed in this framework.

The NA48 experiment is not the first to measure the slope of the scalar form
factor. The first results were obtained in the seventies. In particular, the
high statistics experiment of Donaldson et al.\cite{Donaldson} had confirmed
the theoretical expectations with a slope of $\lambda_0=0.019(4)$. More recent
experiments, however, came up with quite different results. In particular,
three years ago, the KTeV collaboration at Fermilab arrived at a
remarkably small scalar slope: $\lambda_0=0.01372(131)$. Analyzing 0.54
million charged kaon decays, ISTRA, on the other hand, obtained a
much higher value: $\lambda_0= 0.0196(12)(6)$.\cite{ISTRA} If the
results of NA48 as well as those of ISTRA were correct, then the
dependence of the form factors on the momentum transfer would have to show
very strong isospin breaking -- that would be extremely interesting in itself.

My conclusion is that the experimental situation calls for clarification. In
particular, an analysis of the charged kaon decays collected by NA48 might
help removing the dust. There are not many places where the Standard Model
fails. Hints at such failures deserve particular attention.\footnote{In the
  meantime, a preliminary analysis of the KLOE data on the scalar $K^0_{\mu3}$
  form factor became available.\cite{Palutan} The result for the slope,
  $\lambda_0=0.0156(26)$, is in excellent agreement with the theoretical
  prediction, thus removing the puzzle $\ldots\,$ at the price of generating a
  new one: KLOE is in conflict with NA48.}

\section*{Acknowledgments}
It is a pleasure to thank Tran Thanh Van for the kind invitation and for warm
hospitality at La Thuile and Brigitte Bloch-Devaux, Irinel Caprini, Gilberto
Colangelo, Stephan D\"urr, Gerhard Ecker, J\"urg Gasser, Karl Jansen, Roland
Kaiser and Helmut Neufeld for informative discussions and correspondence.

\end{document}